\documentclass[preprint,showpacs,prb,preprintnumbers,amsmath,amssymb,superscriptaddress]{revtex4}

\usepackage{graphicx}%
\usepackage{dcolumn}
\usepackage{amsmath}

\usepackage{color}

\makeatletter
\def\btt#1{\texttt{\@backslashchar#1}}%
\DeclareRobustCommand\bblash{\btt{\@backslashchar}}%
\makeatother

\begin{document}

%\preprint{HEP/123-qed}

\title[Short Title]{Visualization of photogenerated metallic inhomogeneity in the insulating phase of Pr$_{\bf 0.7}$Ca$_{\bf 0.3}$MnO$_{\bf 3}$ thin film using spatiotemporal terahertz radiation imaging}

\author{Noriaki Kida}\email{kida@k.u-tokyo.ac.jp}\affiliation{Department of Advanced Materials Science, The University of Tokyo, 5-1-5 Kashiwa-no-ha, Kashiwa, Chiba 277-8561, Japan}

\author{Masayoshi Tonouchi}\affiliation{Institute of Laser Engineering, Osaka University, 2-6 Yamadaoka, Suita, Osaka 565-0871, Japan}

\date{\today}

\begin{abstract}
Using a spatiotemporal terahertz radiation imaging technique, we visualize photogenerated metallic inhomogeneity in the charge-ordered insulating phase of Pr$_{0.7}$Ca$_{0.3}$MnO$_3$ thin film. We reveal that photogenerated metallic regions with micrometer length scales within the charge-ordered insulating matrix are created on subpicoseond time scales. Such an inhomogeneity becomes homogeneous when the laser power and bias voltage are increased. The observed photogenerated metallic regions would act as nucleus of the insulator-metal transition observed in Pr$_{0.7}$Ca$_{0.3}$MnO$_3$.
\end{abstract}


\maketitle %% NULL FUNCTION WITH LATEX 2e

There is great interest for colossal magnetoresistance (CMR) effect observed in perovskite manganites, in which gigantic change of the resistivity up to 10 orders of magnitude can be induced by applying the magnetic field $H$.\cite{YTokuraRev} Noticeably, a dramatic change of the resistivity in response to $H$, which is accompanied by the insulator-metal transition, strongly enhances near the bicritical point, where the ferromagnetic metallic and antiferromagnetic charge-ordered (CO) insulating phases coexist or compete with each other.\cite{YTokuraRev,EDagotto} The phase coexistence or phase separation, which is  characterized by the existence of the intrinsic inhomogeneity, has been recognized to understand extraordinary properties of strongly correlated electron systems, including high-temperature copper oxide superconductors.\cite{EDagotto} In the case of CMR manganites, such a phase coexistence with characteristic length scales ranging from nanometers to micrometers can be detected by various static experimental techniques such as electron,\cite{MUehara} scanning tunneling,\cite{MFath} magnetic force,\cite{HOhshima} scanning superconducting quantum interference device (SQUID),\cite{YOkimoto} and magneto-optical microscopies.\cite{MTokunaga} For example, ferromagnetic domains with hundreds of micrometers in a ferromagnetic relaxor, Cr-doped CO manganite Pr$_{0.5}$Ca$_{0.5}$MnO$_3$, were observed by a scanning SQUID microscope.\cite{YOkimoto}. The photogenerated metallic patches out of the CO insulating region were also detected in Pr$_{0.7}$Ca$_{0.3}$MnO$_3$.\cite{MFiebig1,MFiebig2} Although ultrafast creation and annihilation processes of the metallic phase in CO manganites were captured by using femtosecond pump-and-probe reflection spectroscopy at visible frequencies\cite{MFiebig4,YOkimoto2} and terahertz radiation,\cite{NKidaTHzRad1,NKidaTHzRad2,NKidaTHzRad3,NKidaTHzRad4,NKidaTHzRad5} there is no report concerning the real-space imaging of the metallic inhomogeneity on picosecond time scales.

Here we adopt the spatiotemporal terahertz radiation imaging technique and report the successful real-space imaging of the inhomogeneity of the subpicosecond photogenerated metallic regions in the CO insulating phase of the typical manganite, Pr$_{0.7}$Ca$_{0.3}$MnO$_3$. The bulk single crystalline sample of Pr$_{0.7}$Ca$_{0.3}$MnO$_3$ exhibits the insulating behavior with the charge ordering temperature about 230 K and becomes the antiferromagnetic insulator at low temperature.\cite{YTokuraRev} Pr$_{0.7}$Ca$_{0.3}$MnO$_3$ is a fascinating compound since CO insulating phase below 230 K can transform to the metallic phase not only by an application of the magnetic field (CMR effect)\cite{YTomioka} but also by other external perturbations such as photoirradiation.\cite{KMiyano} Previously, we have succeeded in detecting the terahertz radiation into free space from a voltage-biased Pr$_{0.7}$Ca$_{0.3}$MnO$_3$ thin film excited by femtosecond laser pulses (Ref. \onlinecite{NKidaTHzRad1}). The terahertz radiation mechanism was picosecond electrical transients or creation of metallic phase within the CO insulating phase. This unique characteristic was used as a probe for the photoinduced insulator-metal transition of this compound (Refs. \onlinecite{NKidaTHzRad2,NKidaTHzRad3,NKidaTHzRad4,NKidaTHzRad5}). On the basis of terahertz radiation characteristics in electric fields at 10 K\cite{NKidaTHzRad4} as well as at 125 K \cite{NKidaTHzRad5}, we inferred that the photogenerated metallic patch within the CO insulating phase reconstructs in response to electric fields. In this work, we directly visualize the distribution of the picosecond electrical transients of Pr$_{0.7}$Ca$_{0.3}$MnO$_3$ thin film at the optically excited area by monitoring the electric field of the radiated terahertz wave.\cite{NKidaTHzRev,MTonouchiRev} On subpicosecond time scales, we find the clear signature of the metallic inhomogeneity even in the robust CO insulating phase at 10 K.

%********************************************************Figure 1
\begin{figure}[t]
\includegraphics[width=0.7\textwidth]{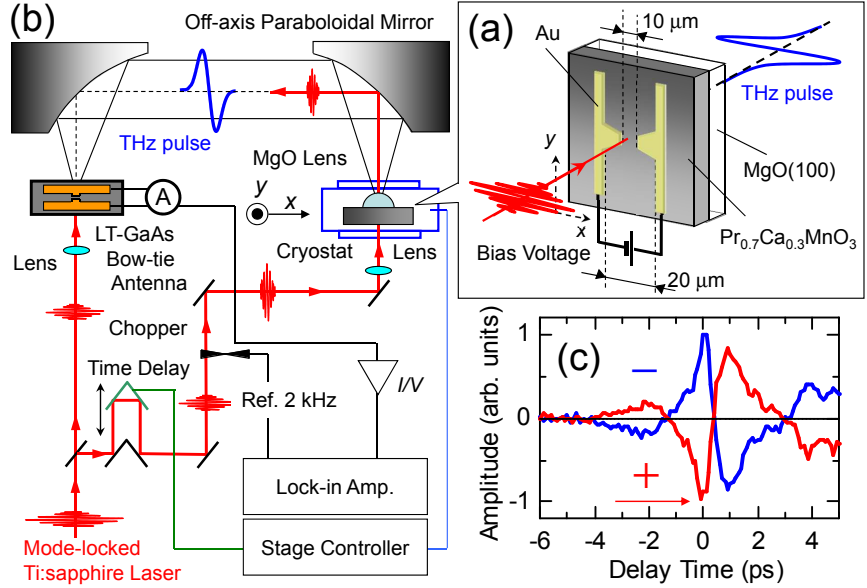}
\caption{(color online) Terahertz radiation experiments of Pr$_{0.7}$Ca$_{0.3}$MnO$_3$ thin film. (a) Scheme of terahertz radiation from voltage-biased Pr$_{0.7}$Ca$_{0.3}$MnO$_3$ thin film excited by femtosecond laser pulses. The dipole antenna made from Au (also acting as an electrode) was patterned on Pr$_{0.7}$Ca$_{0.3}$MnO$_3$ thin film. (b) Schematic illustration of our experimental setup for scanning terahertz radiation imaging experiments. We used the photoconducting sampling technique to directly measure the wave form of terahertz pulse radiated from the Pr$_{0.7}$Ca$_{0.3}$MnO$_3$ thin film. The detector was the bow-tie antenna on the low-temperature-grown GaAs. (c) Propagated terahertz radiation in time-domain at the gap of the antenna ($x=100$ $\mu$m and $y=200$ $\mu$m)  in $V_{\rm bias}$ of $\pm8$ V and $P_{\rm power}$ of 50 mW, measured at 10 K. To obtain the spatiotemporal terahertz radiation image shown in Figs. \ref{fig2}(a), \ref{fig2}(b), and \ref{fig2}(c), we fixed the delay-stage at the maximum peak amplitude of terahertz radiation [marked by an arrow in (c)] and moved the sample along the $x$- and $y$-axes by using stepping motor.
}
\label{fig1}
\end{figure}
%***********************************************************************

Thin film of a Pr$_{0.7}$Ca$_{0.3}$MnO$_3$ on a MgO(100) substrate was prepared by a pulsed laser deposition technique with the use of KrF laser (center wavelength of 193 nm, pulse width of 10 ns, and repetition rate of 5 Hz) as a light source. The substrate temperature and the oxygen base pressure were 770$^\circ$C and 220 Pa, respectively.\cite{NKidaTHzTDS3} The obtained thin film was $a$-axis oriented without any impurity peaks, as characterized by room temperature x-ray diffraction measurements. We patterned the dipole antenna structure made of Au on a Pr$_{0.7}$Ca$_{0.3}$MnO$_3$ thin film by a sputtering technique. The gap distance of the dipole antenna and the distance between two lines were set to 10 $\mu$m and 20 $\mu$m, respectively, as schematically shown in Fig. \ref{fig1}(a). To directly monitor the radiated terahertz pulse in time-domain, we adopted the photoswitching sampling technique in transmission geometry with the use of the low-temperature-grown GaAs (LT-GaAs) coupled with the bow-tie antenna as a detector, as schematically shown in Fig. \ref{fig1}(b).\cite{MTonouchiRev,NKidaTHzRev} The femtosecond laser pulses delivered from a mode-locked Ti:sapphire laser (center wavelength of 800 nm, pulse width of 100 fs, and repetition rate of 82 MHz) were divided by the beam splitter. One was used as the trigger pulse and introduced to the LT-GaAs detector. Another was used as the pump pulse and introduced to the photoswitching device made on Pr$_{0.7}$Ca$_{0.3}$MnO$_3$ thin film [depicted in Fig. \ref{fig1}(a)] after the appropriate time delay. The sample was attached to the holder in the continuous-type cryostat and kept the temperature of 10 K stable. The bias voltage was applied during terahertz radiation experiments. We attached the MgO hemispherical lens on the back side of MgO(100) substrate to enhance terahertz radiation efficiency by reducing the impedance mismatch.\cite{MTonouchi2} The radiated terahertz pulse was collimated by a pair of the off-axis paraboloidal mirrors and introduced to the LT-GaAs detector. The spatiotemporal terahertz radiation image was obtained by fixing the time delay at the maximum absolute peak amplitude $E_{\rm THz}$ of terahertz radiation [as indicated by an arrow in Fig. \ref{fig1}(c)] and by subsequently moving the sample along $x$- and $y$-axes [Fig. \ref{fig1}(a)]. Details of our experimental setup for the measurements of the spatiotemporal terahertz radiation imaging can be found in Ref. \onlinecite{MTonouchi}.

We show in Fig. \ref{fig1}(c) typical example of the radiated terahertz wave forms in time-domain, measured at 10 K. The bias voltage $V_{\rm bias}$ and the laser power $P_{\rm power}$ were fixed to be $\pm8$ V and  of 50 mW, respectively. Within the framework of the current surge model, which were previously confirmed to be hold in the case of Pr$_{0.7}$Ca$_{0.3}$MnO$_3$ (Refs. \onlinecite{NKidaTHzRad4,NKidaTHzRad5}), $E_{\rm THz}$ linearly depends on the electric field $E_{\rm bias}$. This is simply given by
\begin{equation}
E_{\rm THz}\propto\frac{\eta_0 \sigma_{\rm photo} E_{\rm bias} }{\eta_0 \sigma_{\rm photo} +1+\sqrt{\epsilon_{\rm s}}},
\label{eqn.1}
\end{equation}
where $\epsilon_{\rm s}$ the relative dielectric constant of Pr$_{0.7}$Ca$_{0.3}$MnO$_3$, $\eta_0$ the impedance of vacuum (377 $\Omega$), and $\sigma_{\rm photo}$ the photoconductance. This relation holds\cite{NKidaTHzRad1,NKidaTHzRad4} unless $E_{\rm bias}$ exceeds the regime, where $E_{\rm THz}$ irreversibly decreases at temperature close to the spin ordering temperature $\sim130$ K.\cite{NKidaTHzRad5} In the present experiment at 10 K, we confirmed the phase reversal of the wave form by the polarity reversal of $V_{\rm bias}$ [Fig. \ref{fig1}(c)]. This clearly indicates that the transient surge current rather than the nonlinear optical effect, is  source of the observed terahertz radiation. Therefore, we used this quantity as a measure of the amount of the photogenerated metallic patches in the CO insulating phase.

%********************************************************Figure 2
\begin{figure}[t]
\includegraphics[width=0.9\textwidth]{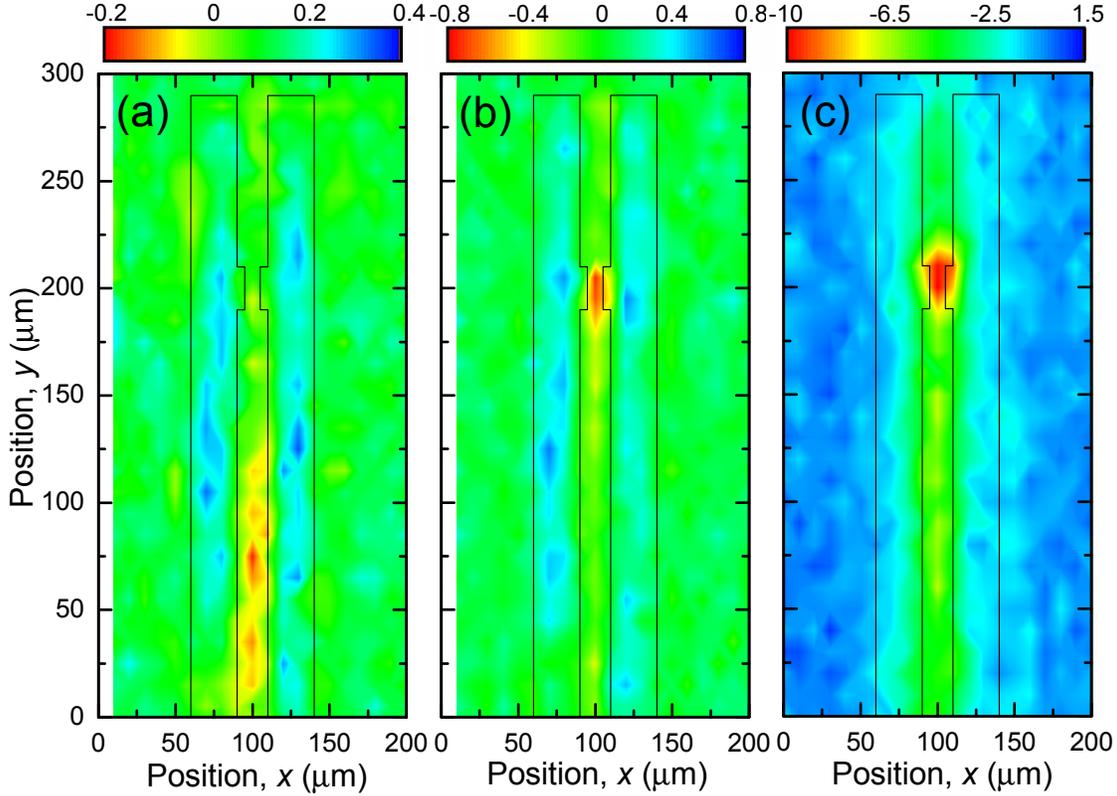}
\caption{(color online) Spatiotemporal terahertz radiation images of Pr$_{0.7}$Ca$_{0.3}$MnO$_3$ thin film, measured at 10 K. The dipole antenna made from Au (indicated by solid lines) were also superimposed on respective images. (a) Terahertz radiation image measured in $V_{\rm bias}$ of 8 V and $P_{\rm power}$ of 50 mW. (b) After removal of external perturbations, terahertz radiation image was taken again in the same condition ($V_{\rm bias}$ of 8 V and $P_{\rm power}$ of 50 mW). (c) Uniform metallic patches were formed when $V_{\rm bias}$ and $P_{\rm power}$ were set to 18 V and 60 mW, respectively. We obtained these terahertz radiation images by measuring $E_{\rm THz}$ at 0 ps [marked by an arrow in Fig. \ref{fig1}(c)] and subsequently by moving the sample along $x$- and $y$-axes [Fig. \ref{fig1}(a)]. Color scale bar of $E_{\rm THz}$ is shown in an upper side of respective figures. Red region indicates the maximum absolute value of $E_{\rm THz}$.}
\label{fig2}
\end{figure}
%***********************************************************************

Figure \ref{fig2}(a) shows the spatiotemporal terahertz radiation image, measured at 10 K. The measurement was performed in $V_{\rm bias}$ of 8 V and $P_{\rm power}$ of 50 mW after the sample was cooled to 10 K in the dark condition. The time needed to take this image was about 18 minutes. The shape of the dipole antenna (indicated by solid lines) was also superimposed on the terahertz radiation image. By considering the shape of the dipole antenna, the electric field $E_{\rm bias}$ between the gap is twice, compared to $E_{\rm bias}$ between electrodes, i.e., $E_{\rm bias}$ of 8 kV/cm at the gap ($d=10$ $\mu$m) and $E_{\rm bias}$ of 4 kV/cm at the position far from the gap ($d=20$ $\mu$m). Thus, according to Eq. (\ref{eqn.1}), $E_{\rm THz}$ at the gap is twice, compared to $E_{\rm THz}$ at the position far from the gap. However, we observed no terahertz radiation around the gap [Fig. \ref{fig2}(a)]. Furthermore, the inhomogeneous distribution of $E_{\rm THz}$ apparently shows up at the position far from the gap. Especially, four regions (colored by red and yellow) with tens of micrometers, are clearly isolated from the background level. 

\begin{figure}[t]
\includegraphics[width=0.9\textwidth]{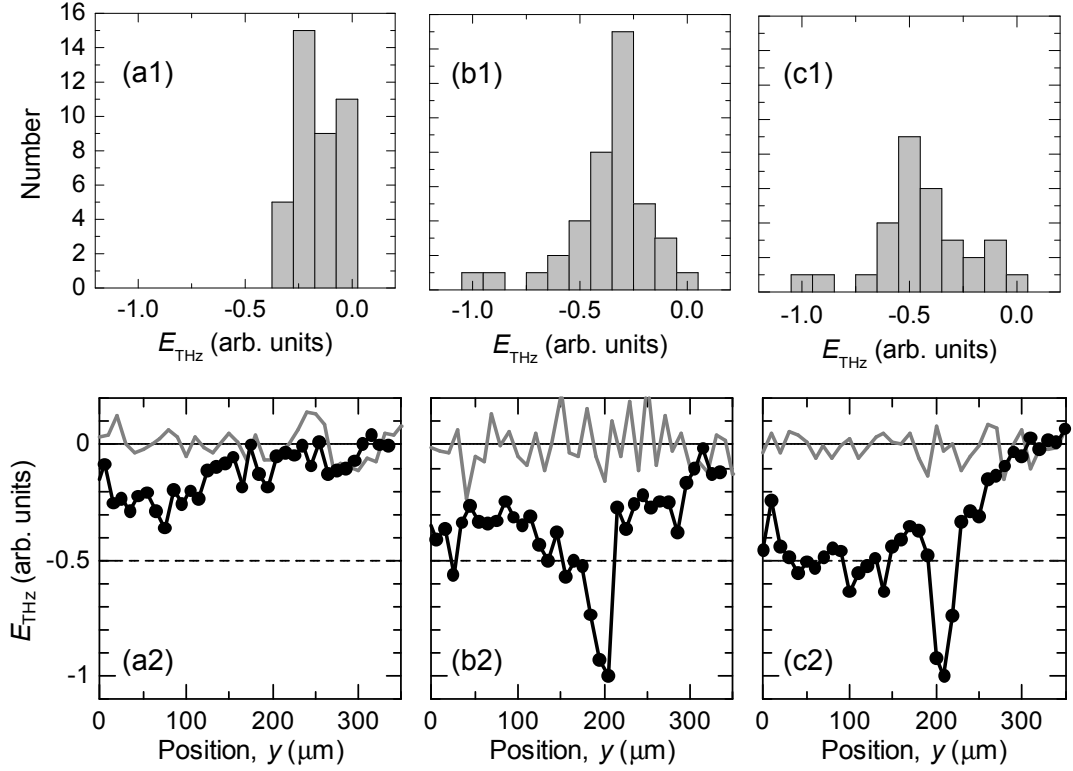}
\caption{Line profile of terahertz radiation along the $y$-axis, measured at 10 K in (a) $V_{\rm bias}$ of 8 V and $P_{\rm power}$ of 50 mW, (b) $V_{\rm bias}$ of 8 V and $P_{\rm power}$ of 50 mW after removal of external perturbations, and (c) $V_{\rm bias}$ of 18 V and $P_{\rm power}$ of 60 mW, as extracted from Figs. \ref{fig2}(a), \ref{fig2}(b), and \ref{fig2}(c), respectively. The circles and gray lines  in the lower panels represent the data at $x=100$ $\mu$m and $x=10$ $\mu$m, respectively. Solid lines are merely the guides to the eye. $E_{\rm THz}$ is normalized by the maximum absolute value of $E_{\rm THz}$, except for $E_{\rm THz}$ in (a2). The dotted line indicates the expected $E_{\rm THz}$ at the position far from the gap of the antenna, which is inferred from the linear relationship of $E_{\rm THz}$ with $E_{\rm bias}$ [see Eq. (\ref{eqn.1})]. Upper panels show the histograms of the respective $E_{\rm THz}$ at $x=100$ $\mu$m.}
\label{fig3}
\end{figure}

The spot diameter of the femtosecond laser pulses was estimated to be about 5 $\mu$m, which is enough to be ruled out the possibility of the non-uniform illumination of the femtosecond laser pulses. One possibility to explain above findings is the presence of the inhomogeneous distribution of $E_{\rm bias}$ due to the imperfection of the patterned antenna structure and/or to the roughness of our sample. To exclude such extrinsic possibilities, we measured the terahertz radiation image while both $V_{\rm bias}$ and $P_{\rm power}$ were increased to 18 V and 60 mW, respectively, at which the uniform metallic patch is formed, as reported in detail in our previous study.\cite{NKidaTHzRad1} Figure \ref{fig2}(c) presents the obtained terahertz radiation image. As we repeatedly reported,\cite{NKidaTHzRad1,NKidaTHzRad4} $E_{\rm THz}$ between the gap linearly changes with $E_{\rm bias}$ unless the persistent insulator-metal transition occurs.\cite{NKidaTHzRad5} To check whether $E_{\rm THz}$ is proportional to $E_{\rm bias}$ or not, we plot in Fig. \ref{fig3}(c2) the line profile of $E_{\rm THz}$ along the $y$-axis at $x=100$ $\mu$m. The data were represented by closed circles, which were normalized by the absolute value of $E_{\rm THz}$ at the gap. The noise level is also shown by a gray line, which is extracted from the line profile of $E_{\rm THz}$ along the $y$-axis at $x=10$ $\mu$m. As can be seen, $E_{\rm THz}$ between electrodes at positions far from the gap is about half of $E_{\rm THz}$ at the gap, as indicated by the dotted line. In order to clearly identify the average value of $E_{\rm THz}$, we present in Fig. \ref{fig3}(c1) the histogram of $E_{\rm THz}$ along the $y$-axis at $x=100$ $\mu$m. In this histogram plot, the number of $E_{\rm THz}$ with a step of 0.1 was counted from the measured terahertz radiation image. As clearly seen in Fig. \ref{fig3}(c1), there are two peak structures centered about $-1$ and $-0.5$, indicating that the linearity of $E_{\rm THz}$ with $E_{\rm bias}$ holds in our present sample. These results [Figs. \ref{fig3}(c1) and \ref{fig3}(c2)] ensure that the observed inhomogeneity shown in Fig. \ref{fig2}(a) cannot be ascribed to the extrinsic factors such as the inhomogeneous distribution of $E_{\rm bias}$.

%*******************************************************Figure 3
%***********************************************************************

The emergent position of the photogenerated metallic regions shown in Fig. \ref{fig2}(a) seems to be arbitrary and is random in real-space since $E_{\rm bias}$ between electrodes is identical. This observation reminds us of the slow relaxation and aging effects, typical characteristics of the phase coexistence, which were frequently observed in CO manganites including Pr$_{0.7}$Ca$_{0.3}$MnO$_3$ (Ref. \onlinecite{AAnane}). From the view point of such a phase coexistence, the photogenerated metallic regions also depend on the external stimulation procedure. Therefore, we removed $V_{\rm bias}$ and $P_{\rm power}$ and subsequently measured the terahertz radiation image in the same condition ($V_{\rm bias}=$ 8 V and $P_{\rm pump}=$ 50 mW). Indeed, the distribution of $E_{\rm THz}$ was dramatically reconstructed, as shown in the terahertz radiation image of Fig. \ref{fig2}(b). This is consistent with the view point of the slow relaxation and aging effects. Noticeably, the stronger $E_{\rm THz}$ shows up around the gap during this procedure. Average value of $E_{\rm THz}$ between electrodes, which is extracted from terahertz radiation image shown in Fig. \ref{fig2}(b), can be estimated to be about $-0.3$ [Fig. \ref{fig3}(b2)]. This value is apparently below the expected value of $E_{\rm THz}$ (dotted line), indicating the violation of the linearity in response to $E_{\rm bias}$. This is also clearly seen in  the histogram plot, revealing the peak about $-0.3$ of $E_{\rm THz}$ [Fig. \ref{fig3}(b1)]. Such a violation of the linearity is more evident for the first terahertz radiation image [Fig. \ref{fig3}(a2)] [In this case, $E_{\rm THz}$ were normalized by the maximum absolute value used in Fig. \ref{fig3}(b2)] and the corresponding histogram plot peaking only about $-0.25$ [Fig. \ref{fig3}(a1)]. Therefore, it is reasonably consider that the observed nonlinear response of $E_{\rm THz}$ as well as appearance of the randomness come from the intrinsic inhomogeneous distribution of the photogenerated metallic patches. Furthermore, the irregular change of the volume fraction of the photogenerated metallic patches was also identified in Figs. \ref{fig2}(a) and \ref{fig2}(b). Therefore, our present study has important implication that there is the metallic inhomogeneity on subpicosecond time scales, even in a robust CO insulating phase at 10 K. With increasing external perturbations, each metallic patch between the gap is merged into the single region and becomes homogeneous, as evidenced by the uniform distribution of terahertz radiation shown in Fig. \ref{fig2}(c).

In summary, we have succeeded in visualizing subpicosecond photogenerated metallic inhomogeneity in the CO insulating phase of the typical CMR manganite, Pr$_{0.7}$Ca$_{0.3}$MnO$_3$ thin film, by mapping out the electric field of terahertz radiation. Such an inhomogeneity is created even on subpicosecond time scales and is highly random with the length scale of tens of micrometers, which could be modified by external perturbations, i.e., bias voltage and photoirradiation. The spatiotemporal terahertz radiation imaging technique used here would yield the complemented information of the electronic inhomogeneity, as widely revealed by the measurements of the static spatially-resolved images of the CMR manganites.\cite{MUehara,MFath,HOhshima,YOkimoto,MTokunaga,MFiebig1,MFiebig2} Furthermore, this technique can be easily combined with a phase sensitive detection. This provides further insights into the nature of exotic properties of strongly correlated electron systems, partly reported in multiferroic BiFeO$_3$ thin films.\cite{KTakahashi,DSRana}

\end{document}